\documentclass{llncs}
\usepackage{llncsdoc}
\usepackage{booktabs}
\usepackage{url}
\usepackage{amssymb}
\usepackage{amsmath}
\usepackage{wrapfig}
\usepackage{enumitem}
\usepackage{graphicx}
\usepackage{epstopdf}
\usepackage{mdwlist}
\usepackage[caption=false]{subfig}
\usepackage{microtype}
\usepackage{color}
\usepackage{capt-of}
\usepackage{sidecap}
\usepackage{algorithm}
\usepackage{eurosym}
\usepackage[misc]{ifsym}
\usepackage{tikz}
\usepackage{rotating}
\usepackage{makecell}
\usetikzlibrary{arrows,automata,backgrounds,decorations,fit,petri,positioning,petri,shapes,calc,patterns,arrows.meta,shadows}
\tikzset{small/.style={level distance=10pt,sibling distance=-2pt,outer ysep=-2pt, label distance=-8pt}}
\tikzset{smaller/.style={level distance=20pt}}
\tikzset{smallish/.style={level distance=20pt,sibling distance=-2pt,outer ysep=-2pt, label distance=-8pt}}
\tikzset{node distance=5mm} 
\tikzset{place/.append style={circle,draw=black,thick,inner sep=0pt,minimum size=4mm,label position=below}} 
\tikzset{transition/.append style={rectangle,draw=black,thick,inner sep=0pt,minimum size=6.8mm}}
\tikzset{tau/.style={transition,fill=black}}  

\urldef{\mailsa}\path|{f.mannhardt,n.tax}@tue.nl|

\newcommand{\Lan}                 {\mathfrak{L}}

\newlength{\hatchspread}
\newlength{\hatchthickness}
\newlength{\hatchshift}
\newcommand{\hatchcolor}{}
\tikzset{hatchspread/.code={\setlength{\hatchspread}{#1}},
	hatchthickness/.code={\setlength{\hatchthickness}{#1}},
	hatchshift/.code={\setlength{\hatchshift}{#1}},
	hatchcolor/.code={\renewcommand{\hatchcolor}{#1}}}
\tikzset{hatchspread=3pt,
	hatchthickness=0.4pt,
	hatchshift=0pt,
	hatchcolor=black}
\pgfdeclarepatternformonly[\hatchspread,\hatchthickness,\hatchshift,\hatchcolor]
{custom north west lines}
{\pgfqpoint{\dimexpr-2\hatchthickness}{\dimexpr-2\hatchthickness}}
{\pgfqpoint{\dimexpr\hatchspread+2\hatchthickness}{\dimexpr\hatchspread+2\hatchthickness}}
{\pgfqpoint{\dimexpr\hatchspread}{\dimexpr\hatchspread}}
{
	\pgfsetlinewidth{\hatchthickness}
	\pgfpathmoveto{\pgfqpoint{0pt}{\dimexpr\hatchspread+\hatchshift}}
	\pgfpathlineto{\pgfqpoint{\dimexpr\hatchspread+0.15pt+\hatchshift}{-0.15pt}}
	\ifdim \hatchshift > 0pt
	\pgfpathmoveto{\pgfqpoint{0pt}{\hatchshift}}
	\pgfpathlineto{\pgfqpoint{\dimexpr0.15pt+\hatchshift}{-0.15pt}}
	\fi
	\pgfsetstrokecolor{\hatchcolor}
	\pgfusepath{stroke}
}

\begin{document}
\title{Unsupervised Event Abstraction using Pattern Abstraction and Local Process Models}

\author{Felix Mannhardt \Letter \and Niek Tax \Letter}

\institute{Eindhoven University of Technology\\
\mailsa
}

\maketitle
\begin{abstract}
Process mining analyzes business processes based on events stored in event logs. However, some recorded events may correspond to activities on a very low level of abstraction. When events are recorded on a too low level of abstraction, process discovery methods tend to generate overgeneralizing process models. Grouping low-level events to higher level activities, i.e., event abstraction, can be used to discover better process models. Existing event abstraction methods are mainly based on common sub-sequences and clustering techniques. In this paper, we propose to first discover local process models and, then, use those models to lift the event log to a higher level of abstraction. Our conjecture is that process models discovered on the obtained high-level event log return process models of higher quality: their fitness and precision scores are more balanced. We show this with preliminary results on several real-life event logs.
\end{abstract}

\keywords{Process Discovery, Unsupervised Learning, Event Abstraction}

\section{Introduction}
Process  mining \cite{Aalst2016} is a fast growing research discipline that concerns the analysis of events that are logged during the execution of a business process. 
Recorded events contain information on what was done, by whom, where, when, etc. 
Such event data is often readily available from business information systems such as ERP, CRM, or, workflow management systems. 
Process discovery, the task of automatically generating a process model that accurately describes the business process based on the event data, plays a central role in the process mining field. 
A variety of process discovery techniques have been developed over the years \cite{Aalst2004,Conforti2016,Leemans2013,Werf2009}, generating process models in different notations, such as Petri nets \cite{Reisig2012}, EPC, and BPMN. 
The degree to which a discovered process model represents the event data from which is discovered is typically expressed in several quality dimensions. Two of such quality dimensions are \emph{fitness}: the amount of behavior in the event log that is allowed by the model and \emph{precision}: the model should not be too general by allowing for much more behavior that was not seen in the event log (i.e., it should not be underfitting).

For successful application of process discovery it is crucial that the events logged in the event log directly correspond to the activities that are recognizable for process stakeholders. In practice, this is not always the case, and there can be an n:m-relation between the recorded events and activities of the process \cite{Baier2013,Gunther2009}. Process models that are generated by process discovery when recorded events and activities do not match have semantics that are unclear to process stakeholders. Moreover, a mismatch between events and activities can cause process discovery techniques to  discover underfitting process models that allow for too much behavior (i.e., process models with low precision) \cite{Tax2016c}.

A recent approach to abstract recorded events to high-level activities \cite{Mannhardt2016} uses \emph{activity patterns} to capture the domain knowledge about the relation between the high-level activities and the low-level recorded events. Each activity pattern is a process model that describes the possible behavior in terms of low-level events that are conjectured to be observed during the execution of a certain high-level activity. However, such domain knowledge might not be available, and when the process contains many activities it becomes a tedious task to model each of them manually. 
Local Process Model (LPM) discovery \cite{Tax2016b,Tax2016a} is a technique to automatically discover frequent patterns of process behavior (i.e., LPMs) from an event log. Each LPM, like an \emph{activity pattern}, is a process model that describes the behavior over only a subset of the event types in the log. 

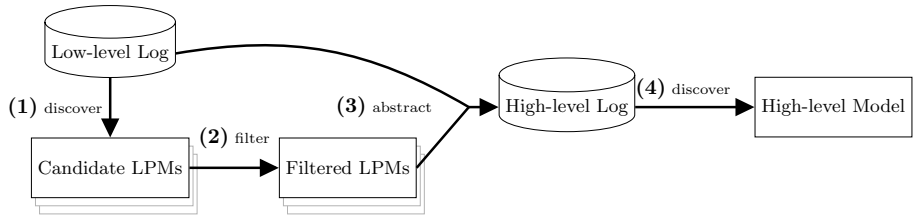
\begin{figure}[t]
\centering
\resizebox{\textwidth}{!}{
\begin{tikzpicture}[
	dir/.tip = {Triangle[scale length=1.5, scale width=1.2]},
	log/.style = {cylinder,shape border rotate=90,aspect=0.25,minimum width=2cm,minimum height=1cm},
	lpm/.style = {minimum width=2cm,minimum height=1cm,double copy shadow={shadow xshift=0.5ex,shadow yshift=-1ex,draw=black!30},fill=white,draw=black},
	model/.style = {minimum width=2cm,minimum height=1cm}
]
\node [draw,log] (llog) {Low-level Log};
\node [draw,lpm, below = 1 of llog] (lpm) {Candidate LPMs};
\node [draw,lpm, right = 1.5 of lpm] (flpm) {Filtered LPMs};
\node [draw,log, below right = -0.25 and 6.5 of llog] (hlog) {High-level Log};
\node [inner sep=0pt,outer sep=0pt,minimum size=0,left = 0.5 of hlog] (placeholder) {};
\node [left = of placeholder,font=\scriptsize] {{\normalsize\textbf{(3)}} abstract};
\node [draw,model, right = 2 of hlog] (hmodel) {High-level Model};
\draw [very thick] 
	(llog) edge[-dir] node[left,font=\scriptsize]{{\normalsize\textbf{(1)}} discover} (lpm)
	(lpm.east) edge[-dir] node[above,yshift=2mm,font=\scriptsize]{{\normalsize\textbf{(2)}} filter} (flpm.west)
	(llog.east) edge[bend left=20] (placeholder)
	(flpm.east) edge (placeholder)
	(placeholder) edge[-dir] (hlog)
	(hlog) edge[-dir] node[sloped,above,xshift=-2mm,font=\scriptsize]{{\normalsize\textbf{(4)}} discover} (hmodel)
;
\end{tikzpicture}}
\caption{An overview of the proposed unsupervised abstraction technique.}%
\label{fig:overview-method}%
\end{figure}

In this paper we explore the application of automatically discovered LPMs to replace the domain knowledge in pattern-based abstraction to form a \emph{novel, completely automated, abstraction technique}. 
Figure~\ref{fig:overview-method} gives an overview of the proposed method. 
Each LPM discovered by the LPM discovery method \cite{Tax2016b,Tax2016a} is assumed to represent a high-level activity. 
Then, we propose a technique to filter the set of LPMs, and use this filtered set of LPMs as activity patterns with the event abstraction method proposed in~\cite{Mannhardt2016} and discover a high-level process model using off-the-shelf discovery methods.
Our proposed, integrated approach aims to \emph{improve the precision} of process models found by process discovery techniques by abstraction of the event log while not sacrificing too much fitness.\looseness=-1

In Section \ref{sec:preliminaries}, we introduce basic concepts and notations. Section \ref{sec:approach} explains how LPM discovery and pattern-based abstraction can be combined to form a fully automated abstraction technique. In Section \ref{sec:results}, we present and discuss some preliminary results, and we conclude the paper and discuss future directions in Section \ref{sec:conclusion}.

\section{Preliminaries}
\label{sec:preliminaries}
In this section we introduce concepts used in later sections of this paper. $X^*$ denotes the set of all sequences over a set $X$ and $\sigma=\langle a_1,a_2,\dots,a_n\rangle$ a sequence of length $n$, with $\sigma(i)=a_i$. $\langle\rangle$ is the empty sequence.

In the context of process logs, we assume the set of all \emph{process activities} $\Sigma$ to be given. An \textbf{event} $e$ in an event log is the occurrence of an activity $e{\in}\Sigma$. We call a sequence of events $\sigma{\in}\Sigma^*$ a \textbf{trace}. An \textbf{event log} $L{\in}\mathbb{N}^{\Sigma^*}$ is a finite multiset of traces. For example, the event log $L=[\langle a,b,c\rangle^2,\langle b,a,c\rangle^3]$ consists of 2 occurrences of trace $\langle a,b,c\rangle$ and three occurrences of trace $\langle b,a,c\rangle$.

A process model notation that is frequently used in the process mining area is the Petri net. A \textbf{labeled Petri net} $N=\langle P,T,F,\ell\rangle$ is a tuple where $P$ is a finite set of \textbf{places}, $T$ is a finite set of \textbf{transitions} such that $P \cap T = \emptyset$,  $F \subseteq (P \times T) \cup (T \times P)$ is a set of directed arcs, called the flow relation, and $\ell:T\nrightarrow \Sigma$ is a labeling function that assigns process activities to transitions. Unlabeled transitions, i.e., $t{\in}T$ with $t{\not}{\in}dom(l)$, are referred to as $\tau$-transitions, or invisible transitions.


The state of a Petri net is defined by its \emph{marking}. 
The marking assigns a finite number of tokens to each place. 
Transition of the Petri net represent activities and can be executed.
The input places of a transition $t \in T$ are all places for which there is a directed edge to the transition: $\{p \in P|(p,t){\in}F\}$. 
The output places of a transition are defined respectively. 
Executing a transition consumes one token from each of its input places and produces one token on each of its output places, i.e., the marking is changed. 
A transition can only be executed when there is at least one token in each of its input places.

Often it is useful to consider a Petri net in combination with an initial marking and a final marking. 
This allows us to define the language, $\Lan(N)$, accepted by Petri net $N$. 
The language of a Petri net is defined by the set of all possible sequences of visible transition labels (i.e., ignoring $\tau$-transitions) that start in the initial marking and end in the final marking. This allows to check whether some  behavior is part of the behavior of the Petri net, i.e., can be replayed on it. 

\begin{figure}[t]
	\centering
	\captionsetup[subfigure]{width=4.5cm}
	\subfloat[]{
		\resizebox{0.5\textwidth}{!}{
			\begin{tikzpicture}
			[node distance=1cm,
			on grid,>=stealth',
			bend angle=20,
			auto,
			every place/.style= {minimum size=5mm},
			caption/.style = {text centered,align=center,pin distance=0.1cm},
			every pin edge/.style = {thick},
			transition/.append style={rectangle,draw=black,thick,inner sep=0pt,minimum size=5mm},
			transitionH/.style={rectangle, thick, fill=black, minimum width=2mm, minimum height=5mm },
			transitionV/.style={rectangle, thick, fill=black, minimum width=5mm, minimum height=2mm }
			]
			\node [place, tokens = 1] (p){};
			\node [transition] (A) [right = of p,pin={[caption]below:{Accepted+Wait}}] {A}
			edge [pre] node[auto] {} (p);
			\node [transitionH] (t1) [above right = 1.1 and 1 of p,label=above:{$\tau_1$}] {}
			edge [pre] node[auto] {} (p);			
			\node [place] (p1) [right = of t1] {}
			edge[pre] node[auto] {} (t1);

			\node [transition] (B) [right = of p1,pin={[caption,pin distance=0.15cm,xshift=-2.5mm]below:{Queued+Awaiting Assignment}}]{B}
			edge[pre] node[auto] {} (p1);
			
			\node [place] (p2) [right = of B] {}
			edge[pre] node[auto] {} (B);

			\node [transitionV] (t2) [above = 0.5 of B,label=above:{$\tau_2$}] {}
			edge[pre] node[auto] {} (p2)
			edge[post] node[auto] {} (p1);

			\node [transitionH] (t3) [right = of p2,label=above:{$\tau_3$}] {}
			edge[pre] node[auto] {} (p2);

			\node [place] (p3) [below = 1.1 of t3] {}
			edge[pre] node[auto] {} (t3)
			edge[pre] node[auto] {} (A);
		
			\node [transition] (C) [right = of p3,pin={[caption]below:{Completed+Closed}}] {C}
			edge[pre] node[auto] {} (p3);
			
			\node [place,pattern=custom north west lines,hatchspread=1.5pt,hatchthickness=0.25pt,hatchcolor=gray] (p4) [right=of C] {}
			edge[pre] node[auto] {}(C);
			\end{tikzpicture}}
		\label{sfig:example_apn}	
	}
	\qquad
	\subfloat[]{
		\centering
		\includegraphics[width=0.4\linewidth]{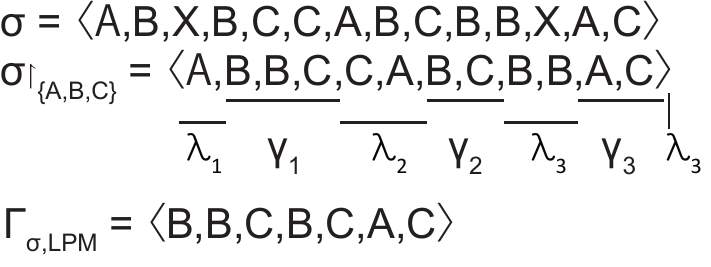}
		\label{sfig:example_alignment}
	}
	\caption{\emph{(a)} Example Petri net $N_1$. \emph{(b)} Trace $\sigma$ of a log $L$ and its segmentation on $N_1$.}
	\label{fig:example}
\end{figure}
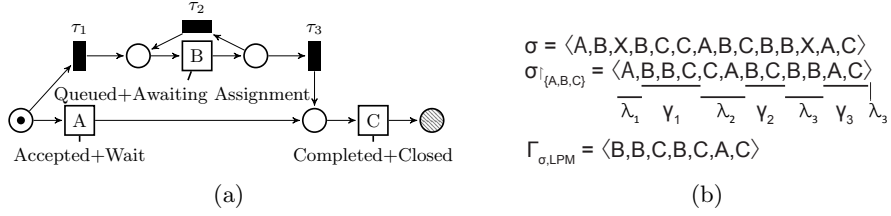

Figure~\ref{sfig:example_apn} shows an example of an accepting Petri net. Circles represent places and rectangles represent transitions. Invisible transitions are depicted as black rectangles. Places that belong to the initial marking contain a token and places belonging to a final marking are marked as $\begin{tikzpicture}
[node distance=1.4cm,
on grid,>=stealth',
bend angle=20,
auto,
every place/.style= {minimum size=0.1mm},
]
\node [place,pattern=custom north west lines,hatchspread=1.5pt,hatchthickness=0.25pt,hatchcolor=gray] {};
\end{tikzpicture}$. The language of this accepting Petri net, with $\Sigma=\{A,B,C\}$ is $\{\langle A,B,C\rangle,\langle B,A,C\rangle,\langle A,B,B,C\rangle,\langle B,B,A,C\rangle,\allowbreak\langle B,A,B,C\rangle,\dots\}$. We refer to \cite{Reisig2012} for a comprehensive introduction of Petri nets. 

\subsection{Local Process Models}
\label{sec:lpm}
LPMs \cite{Tax2016a} are process models that describe the behavior seen in the event log only partially, focusing on frequently observed behavior. 
Typically, LPMs describe the behavior of only up to 5 activities. 
LPMs can be represented in any process modeling notation, such as BPMN, UML, or EPC. Here we use Petri nets to represent LPMs. 
A technique to generate a ranked collection of LPMs through iterative expansion of candidate process models is proposed in \cite{Tax2016a}. 
The search space of process models is fixed, depending on the event log. 
We define $\mathit{LPMS}(L)$ as the set of possible LPMs that can be constructed for given event log $L$. 
We refer the reader to \cite{Tax2016a} for a detailed description of search space $\mathit{LPMS}(L)$.

To evaluate a given LPM on a given event log $L$, its traces $\sigma{\in}L$ are first projected on the set of activities $\Sigma$ in the LPM, i.e., $\sigma'{=}\sigma{\upharpoonright}_{\Sigma}$. 
The projected trace $\sigma'$ is then segmented into $\gamma$-segments that fit the behavior of the LPM and $\lambda$-segments that do not fit the behavior of the LPM, i.e., $\sigma'{=}\lambda_1 \gamma_1 \lambda_2 \gamma_2 \cdots \lambda_n \gamma_n \lambda_{n+1}$ such that $\gamma_i{\in}\Lan(\mathit{LPM})$ and $\lambda_i{\not\in}\Lan(\mathit{LPM})$. 
We define $\Gamma_{\sigma,LPM}$ to be a function that projects trace $\sigma$ on the LPM activities and obtains its subsequences that fit the LPM, i.e., $\Gamma_{\sigma,LPM}=\gamma_1\gamma_2\dots\gamma_n$.

Let our LPM $N_1$ under evaluation be the Petri net of Figure~\ref{sfig:example_apn} and let $\sigma=\langle A,B,X,B,C,C,A,B,C,B,B,X,A,C\rangle$ be an example trace. Function $\mathit{act}(\mathit{LPM})$ obtains the set of process activities in the LPM, e.g. $\mathit{act}(N_1)=\{A,B,C\}$. 
Projection on the activities of the LPM gives $\sigma{\upharpoonright}_{\mathit{act}(N_1)}=\langle A,B,B,C,C\allowbreak,A,B,C,\allowbreak B,B,A,C\rangle$. 
Figure~\ref{sfig:example_alignment} shows the segmentation of the projected traces on the LPM, leading to $\Gamma_{\sigma,LPM}=\langle B,B,C,B,C,A,C\rangle$. 
The segmentation starts with a non-fitting segment $\lambda_1=\langle A\rangle$, followed by a fitting segment $\gamma_1{=}\langle B,B,C\rangle$, which completes one run through the model from initial to final marking. 
The second event $C$ in $\sigma$ cannot be replayed on $LPM$, since it only allows for one $C$ and $\gamma_1$ already contains a $C$. This results in a non-fitting segment $\lambda_2{=}\langle C,A\rangle$. $\gamma_2{=}\langle B,C\rangle$ again represents a run through the model from initial to final marking, and $\lambda_3{=}\langle B,B\rangle$ does not fit the LPM. $\gamma_3{=}\langle A,C\rangle$ again represents a run though the model, and we end with a empty non-fitting segment $\lambda_4$. We lift segmentation function $\Gamma$ to event logs, $\Gamma_{L,LPM}{=}\{\Gamma_{\sigma,LPM}|\sigma{\in}L\}$. An alignment-based \cite{Aalst.2012} implementation of $\Gamma$, as well as a method to rank and select LPMs based on their support, i.e., the number of events in $\Gamma_{L,LPM}$, is described in \cite{Tax2016a}.

\subsection{Pattern-based Abstraction}

\begin{figure}[t]
\centering
\resizebox{\textwidth}{!}{
\begin{tikzpicture}[
	dir/.tip = {Triangle[scale length=1.5, scale width=1.2]},
	log/.style = {cylinder,shape border rotate=90,aspect=0.25,minimum width=2cm,minimum height=1cm},
	lpm/.style = {minimum width=2cm,minimum height=1cm,double copy shadow={shadow xshift=0.5ex,shadow yshift=-1ex,draw=black!30},fill=white,draw=black},
	model/.style = {minimum width=2cm,minimum height=1cm}
]
\node [draw,log] (llog) {Event Log};
\node [draw,lpm, below = of llog] (patterns) {Activity Patterns};
\node [draw,log, right = 4 of llog] (alignedlog) {Aligned Log};
\node [draw,model, below = of alignedlog] (abstractionmodel) {Abstraction Model};
\node [draw,log, right = 4 of alignedlog] (abstractedlog) {Abstracted Log};
\draw [very thick] 
	(llog) edge[-dir] (alignedlog)
	(patterns) edge[-dir] node[above,font=\scriptsize,text width=3cm]{Build Abstraction Model} (abstractionmodel) 
	(abstractionmodel) edge[-dir] node[left,font=\scriptsize]{Align Log} (alignedlog)
	(alignedlog) edge[-dir] (abstractedlog)
	(abstractionmodel) edge[-dir,bend right=15] node[below,font=\scriptsize]{Abstract Log}  (abstractedlog)
;
\end{tikzpicture}}
\caption{An overview of the pattern-based abstraction method described in~\cite{Mannhardt2016}. These steps jointly form step 3 of Figure~\ref{fig:overview-method}."
}%
\label{fig:abstraction}%
\end{figure}
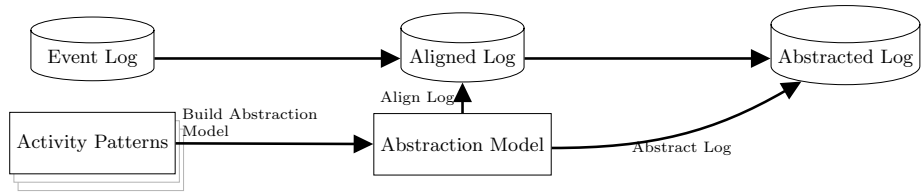

An overview of the pattern-based abstraction method proposed in~\cite{Mannhardt2016} is depicted in Figure~\ref{fig:abstraction}. The input is an event log and a set of activity patterns. Each activity pattern is a process model that represents the behavior expected for one execution of a high-level activity. 
Moreover, a mapping from activities to life-cycle transitions of the high-level activity is required, i.e., $\theta: T \not\rightarrow LT$ with, e.g., $LT = \{ start, complete \}$. Mapping $\theta$ allows to obtain information on when activities started and when they were completed, which some process discovery algorithms, such as \cite{Leemans2015}, are able to leverage.

Activity patterns are composed to an abstraction model and, then, an alignment technique \cite{Aalst.2012} is used to obtain a high-level event log. Through the use of the alignment technique, we can capture approximate executions of activity patterns. Then, we use the alignment information to create a corresponding high-level event log by only retaining those events that were aligned to activities $t \in T$ that are mapped to high-level activities, i.e., $t \in dom(\theta)$.
Activity patterns may use any kind of process models with clear semantics, thus, the abstraction method also works with LPMs.
For example, assume that LPM $N_1$ represents the behavior expected for some high-level activity. Mapping function $\theta$ could be defined such that the transition corresponding to $A$ and transition $\tau_1$ are mapped to \emph{start} and the transition corresponding to $C$ is mapped to \emph{complete}. It is possible to automatically obtain such mapping function $\theta$, e.g., by mapping the source activities to the start and the and the sink activities to the complete life-cycle transitions. 
\section{Unsupervised Abstraction Technique}
\label{sec:approach}

We use discovered LPMs to replace the domain knowledge originally used in the pattern-based event abstraction method. Our proposed method discovers a high-level process model in the following four steps shown graphically in Fig.~\ref{fig:overview-method}.
\begin{enumerate}
	\item We discover a \textbf{fixed number of candidate LPMs} based on the ranking proposed in~\cite{Tax2016a}. The LPM ranking proposed in \cite{Tax2016a} is based on support (i.e., frequency). When setting the number of LPMs to use for abstraction to $k$, we select the top $k$ LPMs of the discovered ranking of LPMs $R$. For event logs with large numbers of activities the original LPM discovery algorithm \cite{Tax2016a} becomes computationally infeasible. For such logs, faster, approximate, techniques for discovering LPMs can be applied \cite{Tax2016b}.
	\item The LPMs in ranking $R$ can overlap in the activities that they describe. For the purpose of event abstraction this can be undesirable, because this results in multiple similar patterns for pattern-based abstraction, making it unclear which low-level event belongs to which high-level activity. Therefore, we introduce a \textbf{diversity score} for the $i$-th LPM of LPM ranking $R$ as: \begin{center}
	$\mathit{div}(R,i){=}
	\left\{
	\begin{array}{ll}
	\max_{j=1}^{i-1}\frac{|\mathit{act}(R(i))\cap\mathit{act}(R(j))|}{|\mathit{act}(R(i))|+|\mathit{act}(R(i))|-|\mathit{act}(R(i))\cap\mathit{act}(R(j))|}  & \mbox{if } i>1, \\
	1 & \mbox{if } i=1.
	\end{array}
	\right.$
\end{center} 
Based on this definition of diversity we introduce a diversity threshold $t_\mathit{div}$ to \textbf{filter out} all LPMs from $R$ where $\mathit{div}(R,i){\le}t_\mathit{div}$. This filter removes LPMs from $R$ when there is a stronger LPM, higher in ranking $R$, that describes (almost) the same set of activities.
	\item We use those LPMs as \textbf{activity patterns} and obtain a high-level event log with the \textbf{abstraction method} described in~\cite{Mannhardt2016}.
	In work~\cite{Mannhardt2016} several composition functions are presented that allow fine grained control of the interaction between activities. In our case, we do not assume any domain knowledge on the process, thus, we limit the choice of composition parameter for our method to: \emph{interleaving} and \emph{parallel}. Interleaving means that any two high-level activities cannot occur at the same time, whereas parallel means that no such restriction is placed.
		\item Based on the high-level event log, we \textbf{discover a high-level process model} using existing methods such as the Inductive Miner (IM)~\cite{Leemans2013}.
\end{enumerate}

For example, take LPM $N_1$ shown in Fig.~\ref{fig:example} and assume that we apply the proposed method to trace $\sigma$. 
We assume that LPM $N_1$ represents the behavior of some high-level activity $H$.
Three executions (i.e., $\gamma$-segments) of the LPM $N_1$ are present in trace $\sigma$, thus, the high-level activity $H$ was executed three times.
When applying pattern-based abstraction, we obtain the high-level trace $\langle H, H, H \rangle$\footnote{The abstraction technique also provides events for the start and complete life-cycle transitions. So far, we only use the complete transition.}.
Some low-level events in $\sigma$ could not be matched to a high-level activity, e.g., the first event of the trace: $A$.
Since we cannot assume that the discovered LPMs represent the entire behavior of the process, we add all those low-level events to the resulting trace.
Therefore, we use the trace $\langle A, H, C, A, H, B, B, H \rangle$, which contains less events than the original trace $\sigma$, for discovery.

Note that our technique does not discover the names of the high-level activities represented by the LPMs. 
However, LPMs can be labeled to its corresponding business activity based on domain knowledge.
\section{Preliminary Results and Discussion}
\label{sec:results}

To be able to deal with the computational complexity of LPM mining for logs with many activities, we discover LPMs using activity clustering based on Markov clustering, as proposed in \cite{Tax2016b}. As proposed in~\cite{Mannhardt.2017}, we expand high-level activities in the discovered process model with the corresponding LPMs to provide a fair comparison based on the low-level event log. The resulting \emph{expanded process model} can be related to the low-level events with existing techniques that determine the models quality in terms of fitness and precision~\cite{Aalst.2012}. We evaluate the quality of the expanded process models in terms of F-score~\cite{DeWeerdt2011}, i.e., the harmonic mean between fitness and precision.

\begin{figure}[t]
\centering
	\scalebox{1}{
	\includegraphics[width=\linewidth]{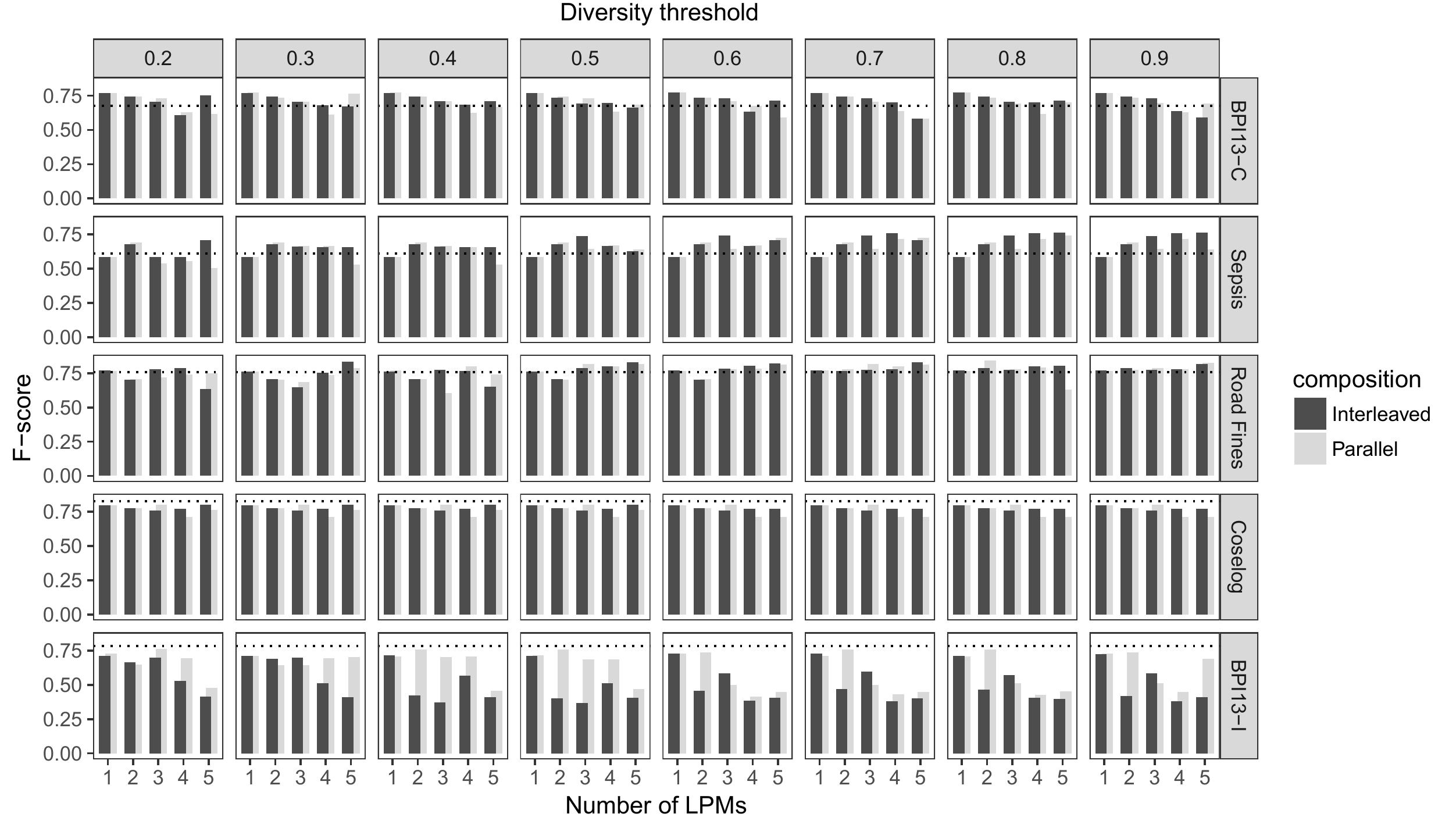}}
	\caption{The F-scores of models discovered with the IM infrequent after applying the proposed method compared to the F-score obtained without the method (dotted line).}
	\label{fig:results}
\end{figure}

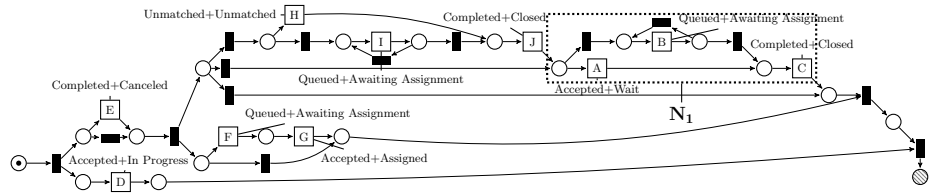
\begin{figure}[t]
\centering
\resizebox{\textwidth}{!}{%
	\begin{tikzpicture}
			[node distance=1cm,
			on grid,>=stealth',
			bend angle=20,
			auto,
			every place/.style= {minimum size=5mm},
			caption/.style = {text centered,align=center,pin distance=0.1cm},
			every pin edge/.style = {thick},
			transition/.append style={rectangle,draw=black,thick,inner sep=0pt,minimum size=5mm},
			transitionH/.style={rectangle, thick, fill=black, minimum width=2mm, minimum height=5mm },
			transitionV/.style={rectangle, thick, fill=black, minimum width=5mm, minimum height=2mm }
			]
			
			\node [place, tokens = 1] (snk){};			
			
			\node [transitionH] (t1) [right = of snk] {}
			edge [pre] node[auto] {} (snk);
			
			\node [place] (p1) [below right = 0.5 and 0.7 of t1] {}
			edge[pre] node[auto] {} (t1);
						
			\node [transition] (A) [right = of p1,pin={[caption,pin distance=0.05cm,xshift=2mm]above:{Accepted+In Progress}}]{D}
			edge[pre] node[auto] {} (p1);
			
			\node [place] (p3) [right = of A] {}
			edge[pre] node[auto] {} (A);			

			\node [place] (p2) [above right = of t1] {}
			edge[pre] node[auto] {} (t1);
			
			\node [transition] (B) [above right = of p2,pin={[caption]above:{Completed+Canceled}}]{E}
			edge[pre] node[auto] {} (p2);
			
			\node [transitionV] (t2) [below = 0.75 of B] {}
			edge [pre] node[auto] {} (p2);

			\node [place] (p4) [below right = of B] {}
			edge[pre] node[auto] {} (B)
			edge[pre] node[auto] {} (t2);			

			\node [transitionH] (t3) [right = of p4] {}
			edge [pre] node[auto] {} (p4);
			
			\node [place] (p5) [below right = of t3] {}
			edge[pre] node[auto] {} (t3);
					
			\node [transitionH] (t4) [right = 1.7 of p5] {}
			edge [pre] node[auto] {} (p5);			

			\node [transition] (C) [above right = of p5,pin={[caption,pin distance=0.1cm]above right:{Queued+Awaiting Assignment}}]{F}
			edge[pre] node[auto] {} (p5);

			\node [place] (p6) [right = of C] {}
			edge[pre] node[auto] {} (C);

			\node [transition] (D) [right = of p6,pin={[caption]below right:{Accepted+Assigned}}]{G}
			edge[pre] node[auto] {} (p6);
			
			\node [place] (p7) [right = of D] {}
			edge[pre] node[auto] {} (D)
			edge[pre,bend left] node[auto] {} (t4);
			
			\node [place] (p8) [above right = 1.8 and 0.75 of t3] {}
			edge[pre] node[auto] {} (t3);			

			\node [transitionH] (t5) [above right = of p8] {}
			edge [pre] node[auto] {} (p8);

			\node [transitionH] (t6) [right = 0.6 of p8] {}
			edge [pre] node[auto] {} (p8);			

			\node [transitionH] (t7) [below right = of p8] {}
			edge [pre] node[auto] {} (p8);			
				
			\node [place] (p9) [right = of t5] {}
			edge[pre] node[auto] {} (t5);
			
			\node [transitionH] (t8) [right = of p9] {}
			edge [pre] node[auto] {} (p9);			

			\node [transition] (E) [above right = of p9,pin={[caption]left:{Unmatched+Unmatched}}]{H}
			edge[pre] node[auto] {} (p9);
			
			\node [place] (p10) [right = of t8] {}
			edge[pre] node[auto] {} (t8);
					
			\node [transition] (F) [right = of p10,pin={[caption,pin distance=0.5cm]below:{Queued+Awaiting Assignment}}]{I}
			edge[pre] node[auto] {} (p10);			

			\node [place] (p11) [right = of F] {}
			edge[pre] node[auto] {} (F);
			
			\node [transitionV] (t9) [below = 0.5 of F] {}
			edge [pre] node[auto] {} (p11)
			edge [post] node[auto] {} (p10);			

			\node [transitionH] (t10) [right = of p11] {}
			edge [pre] node[auto] {} (p11);

			\node [place] (p12) [right = of t10] {}
			edge[pre] node[auto] {} (t10)
			edge[pre,bend right=10] node[auto] {} (E);
			
			\node [transition] (G) [right = of p12,pin={[caption,xshift=-10mm]above:{Completed+Closed}}]{J}
			edge[pre] node[auto] {} (p12);			

			\node [place] (p13) [below right = of G] {}
			edge[pre] node[auto] {} (G)
			edge[pre] node[auto] {} (t6);
			
			\node [transitionH] (t11) [above right = of p13] {}
			edge [pre] node[auto] {} (p13);
						
			\node [transition] (H) [right = of p13,pin={[caption,pin distance=0.05cm]below:{Accepted+Wait}}]{A}
			edge[pre] node[auto] {} (p13);
			
			\node [place] (p14) [right = of t11] {}
			edge[pre] node[auto] {} (t11);

			\node [transition] (J) [right = of p14,pin={[caption]above right:{Queued+Awaiting Assignment}}]{B}
			edge[pre] node[auto] {} (p14);

			\node [place] (p15) [right = of J] {}
			edge[pre] node[auto] {} (J);

			\node [transitionV] (t12) [above = 0.5 of J] {}
			edge [pre] node[auto] {} (p15)
			edge [post] node[auto] {} (p14);
			
			\node [transitionH] (t13) [right = of p15] {}
			edge [pre] node[auto] {} (p15);
						
			\node [place] (p16) [below right = of t13] {}
			edge[pre] node[auto] {} (t13)
			edge[pre] node[auto] {} (H);

			\node [transition] (K) [right = of p16,pin={[caption]above:{Completed+Closed}}]{C}
			edge[pre] node[auto] {} (p16);
			
			\node [place] (p17) [below right = of K] {}
			edge[pre] node[auto] {} (K)
			edge[pre] node[auto] {} (t7);			

			\node [transitionH] (t14) [right = of p17] {}
			edge [pre] node[auto] {} (p17)
			edge [pre,bend left=9] node[auto] {} (p7);

			\node [place] (p18) [below right = of t14] {}
			edge[pre] node[auto] {} (t14);
			
			\node [transitionH] (t15) [below right = of p18] {}
			edge [pre] node[auto] {} (p18)
			edge [pre,bend left=2] node[auto] {} (p3);

			\node [place,place,pattern=custom north west lines,hatchspread=1.5pt,hatchthickness=0.25pt,hatchcolor=gray] (snk) [below = 0.75 of t15] {}
			edge[pre] node[auto] {} (t15);

			\node[draw,dotted,ultra thick,fit=(p13) (K) (t12),pin={[caption,pin distance=0.5cm,font=\Large]below:{$\mathbf{N_1}$}}] (n1) {};			
			\end{tikzpicture}}
	\caption{A Petri net discovered for the BPI13-C event log with parallel composition.}
	\label{fig:result-model}
\end{figure}

Figure~\ref{fig:results} shows the F-score of the process models discovered with the IM infrequent~\cite{Leemans2013} process discovery algorithm with 20\% noise filtering. Horizontally it shows the results for different LPM \emph{diversity thresholds} (0.2 to 0.9). Vertically it shows the results for five different event logs: the BPI challenge 2013 incidents (I) and closed problems (C) log\cite{Steeman.2013}, the CoSeLoG receipt phase log~\cite{Buijs.2014}, the road fines log~\cite{Leoni.2015}, and the sepsis log~\cite{Mannhardt.2016}. 
The results are shown for 1 to 5 LPMs used in abstraction and parallel and interleaving composition.
The dotted line indicates the F-score of the process model discovered from the original event log, i.e., without abstraction.

Figure~\ref{fig:result-model} shows an expanded process model that is discovered for the BPI13-C log based on the proposed method. 
The LPM $N_1$ was used as activity pattern and is part of the process model. Thus, the process model can hierarchically decomposed into sub-processes based on the activity patterns. 
Its precision score is improved from 0.53 to 0.86 at the expense of fitness, which drops from 0.84 to 0.65. The F-score improved from 0.65 to 0.74.

In our preliminary results we found that for three out of five event logs the F-score of the process model can be improved by abstracting the event log prior to process discovery.
Furthermore, it seems that abstracting with parallel composition does only improve the process model over abstraction with interleaving composition in one case, which is beneficial since the interleaving composition is computationally less expensive.
The optimal number of LPMs used for abstraction differs between event logs, with the optimal number of LPMs being 5 for the sepsis log and 1 for the BPI13-C log. To make an abstraction technique based on LPM discovery and pattern-based abstraction fully automated, further experimentation would be needed. We need to analyze whether the optimal number of LPMs depends on properties of the event log and for which logs the good results can be expected.

\section{Conclusions and Future Work}
\label{sec:conclusion}
In this paper we have described a technique to abstract an event log to a higher-level event log using Local Process Model (LPM) discovery and pattern-based abstraction. 
We have shown on five real life event logs that the abstraction approach applied prior to process discovery can result in more precise process models.

We found that (1) the number of LPMs that should be used for abstraction, (2) the diversity threshold, and (3) the composition method which result in good process models being discovered are very dependent on the event log on which the technique is applied. 
In future work, we want to investigate this interplay between event log properties and the parameters of the abstraction approach that are needed to discover process models that strike a good balance between precision and fitness.
We aim to apply this insight in the relation between parameters settings of the abstraction technique and properties of the event log in a technique that can abstract an event log completely automatically, where no manual parameter setting is needed.

\bibliographystyle{splncs03}
\bibliography{bibliography}
\end{document}